\documentstyle[aps,twocolumn]{revtex}
\input{epsf}
\begin{document}
\title{{Nonlinearity effects in the kicked oscillator}}

\vskip4mm                      

\author{Roberto Artuso$^{a,b,c, \dag}$ and
Laura Rebuzzini$^{a}$}
\address{$^{(a)}$Centre for Nonlinear and Complex Systems and
Dipartimento di Scienze Chimiche, Fisiche e Matematiche, Universit\`a 
dell'Insubria, Via Valleggio 11,
22100 Como, Italy}
\address{$^{(b)}$Istituto Nazionale di Fisica della Materia, 
Unit\`a di Como, Via Valleggio 11, 22100 Como, Italy}
\address{$^{(c)}$Istituto Nazionale di Fisica Nucleare, Sezione di Milano,
Via Celoria 16, 20133 Milano, Italy} 
\address{$^{(\dag)}${ email address; roberto.artuso@uninsubria.it}}

\maketitle

\begin{abstract}

{The quantum kicked oscillator is known to display a remarkable
richness of dynamical behaviour, from ballistic spreading to dynamical 
localization. Here we investigate the effects of a Gross Pitaevskii
nonlinearity on quantum motion, and provide evidence that the 
qualitative features depend strongly on the parameters of the system.

PACS numbers:05.45.-a 

}
\end{abstract}

\pacs{PACS numbers:05.45.-a }

\narrowtext


The dynamical behavior of quantized area-preserving maps has proved to 
be one the most relevant field in the discipline of quantum chaos (see
\cite{QuCh}): in particular the discovery of {\em quantum dynamical
localization}\cite{CCFI}, namely the quantal suppression of 
classical deterministic diffusion, has provoked a vast amount of 
theoretical and experimental work. 
The paradigmatic example in this field is the quantum kicked rotator 
(see \cite{Iz}), obtained upon quantization of the classical standard 
map\cite{BVC}. From a classical point of view the system is of KAM 
nature: for small values of the stochasticity parameter global 
transport is inhibited by invariant curves: once the last invariant 
torus is destroyed, transport properties abruptly change, and, in 
typical situations of strong chaos, there is a diffusive spreading in 
momentum\cite{BVC,DK}, which characterizes also the quantum motion for 
times shorter than the break-time $t_b$, where quantum localization 
regime sets in and the momentum spreading is suppressed\cite{note1}.
We remark that this picture is valid for {\em generic} values of the 
effective Planck's constant: quantum resonant motion, characterized by 
ballistic spreading, appears when $\hbar$ assumes rational 
values\cite{CCFI,resIS}.

Another example of a quantum system originating from a two dimensional 
area preserving map is the kicked harmonic oscillator (see 
\cite{BRZ,BR,SS} and references therein): the classical hamiltonian is 
\begin{equation}
{\cal H}(p,x,t)\,=\,\frac{1}{2 m_0}p^2\,+\,\frac{m_0}{2} \omega_0^2 x^2 
\, + \, \varepsilon \cos (k_0x ) \delta_{T_0}(t)
\label{classH}
\end{equation}
where the time dependence is through the periodic delta 
function
\[
\delta_{T_0}(t)\,=\,\sum_{m=-\infty}^{\infty}\, \delta(t-T_0m)
\]
By rescaling variables $\tilde{x}=k_0x$ and $\tilde{t}=\omega_0 t$ we 
realize that dynamics is only dependent on the parameters
\begin{equation}
{\cal K}\,=\,\frac{\varepsilon k_0^2}{m_0 \omega_0} \qquad \quad 
T\,=\,T_0 \omega_0
\label{classPAR}
\end{equation}
In particular the {\em resonant} case ($T=2 \pi p/q$) is characterized 
by the presence of a stochastic web (for arbitrarily small values of 
${\cal K}$) supporting unbounded transport\cite{c-Zas,KAMnote}, while in the 
non-resonant case a threshold ${\cal K}(E_0)$ exists below which 
unbounded transport is not sustained\cite{c-Zas2}.

The kicked harmonic oscillator has been proposed as a model of 
different physical phenomena: from electronic transport in 
semiconductor superlattices\cite{From}, to ion traps\cite{GCZ}. In the 
latter case the harmonic potential is representative of the ion trap, 
while the kicking term arises from a time periodic standing wave laser 
field. Obviously such examples require a proper quantum mechanical 
treatment of the Hamiltonian (\ref{classH}), the corresponding 
Schroedinger equation being (once expressed in dimensionless variables
$\tilde{t}=\omega_0 t$ and $\tilde{x}=\sqrt{m_0 \omega_0/\hbar}x$)
\begin{equation}
i \frac{\partial}{\partial \tilde{t}} \psi\,=\, \left( -\frac{1}{2} 
\frac{\partial^2}{\partial \tilde{x}^2} \, +\, \frac{1}{2} \tilde{x}^2 \, +
\, \sigma \cos(\xi \tilde{x}) \delta_{T}(\tilde{t}) \right)\psi
\label{quanS}
\end{equation} 
so that quantum dynamics depends upon three parameters
\begin{equation}
\sigma=\frac{\varepsilon}{\hbar}\qquad \quad \xi=k_0 
\sqrt{\frac{\hbar}{m_0 \omega_0}}\qquad \quad T=T_0 \omega_0
\label{qpar}
\end{equation}
Once again the behavior is quite sensitive to number theoretic 
properties of $T$: in particular the {\em crystal} cases\cite{BR}
$T=2 \pi/q$ with $q \in \{1,2,3,4,6\}$ admit a one-parameter group of 
commuting generalized translations commuting with the hamiltonian (exceptional parameter 
values\cite{BR} may also lead to two-parameter groups): the 
corresponding dynamical behavior is diffusive (or ballistic in the 
exceptional cases). We remark that the $q=4$ case corresponds to the 
(symmetric) kicked Harper model\cite{kHm}. Outside resonant parameter 
values there are indication of a localization-delocalization transition 
for resonant-non crystal cases\cite{SS}, while the simulations reported 
in \cite{BR} for non-resonant cases suggest dynamical 
localization\cite{obs-plane} (however we observed a delocalization 
transition in the irrational case too).

Recently it has been suggested\cite{GJDCZ} how the widespread interest 
and experimental activity in Bose-Einstein condensation\cite{bec} makes it 
natural to study the effect on Gross-Pitaevskii nonlinearities\cite{tn} on the 
kicked oscillator (thus turning it into a model of a trapped condensate
under a laser field in the spirit of \cite{GCZ}).
The Gross-Pitaevskii nonlinear correction to the Schroedinger equation 
is of the form $u |\psi|^2 \psi$, where the coefficient $u$ is of the 
same sign of the scattering length\cite{obsGP} (we will mainly deal with a positive 
$u$ in what follows): using the same rescaling in dimensionless variables 
mentioned in the quantum case the equation reads
\begin{equation}
i\frac{\partial}{\partial 
\tilde{t}}\psi \,=\,\left(-\frac{1}{2}\frac{\partial}{\partial \tilde{x}^2} 
+\frac{1}{2}\tilde{x}^2 + \sigma \cos(\xi \tilde{x})\delta_T(\tilde{t}) 
+ v |\psi|^2 \right) \psi
\label{GPe}
\end{equation}
where now
\begin{equation}
v=\frac{u}{\hbar}\sqrt{\frac{m_0}{\hbar \omega_0}}
\label{GPpar}
\end{equation}
Even if the cubic nonlinearity acts like an effective {\em repulsive} 
potential, the main observation in \cite{GJDCZ}, as regards the 
dynamical effect of the Gross-Pitaevskii nonlinearity in a crystal $q=6$ 
case, has been its tendency to oppose quantum spreading: it was 
suggested that 
this is due to a breakup of quantum symmetries for nonzero 
nonlinearity.
Before presenting the results of our simulations we have to mention 
that nonlinearity effects have also been considered a few years ago for
the kicked rotator\cite{zen,dN} (for a cubic nonlinearity of 
opposite sign). Here the scenario is quite different: when the 
nonlinearity is absent the system exhibits quantum dynamical 
localization: a sufficiently strong nonlinearity may then induce 
chaotic transitions between localized modes, leading to (subdiffusive) 
delocalization.

\begin{figure}[h]
\label{fig-1}
\centerline{\epsfxsize=5.5cm \epsfbox{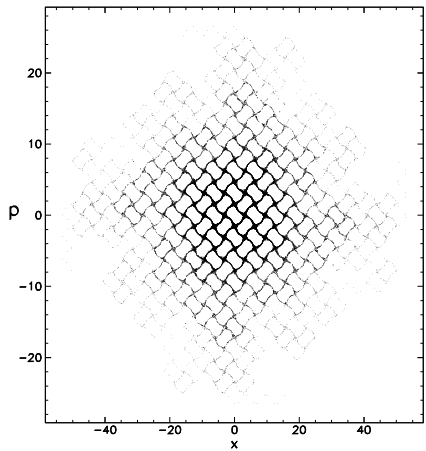}}
\caption{Classical transport along the stochastic web.}
\end{figure}
To investigate the effect of the nonlinearity, we studied the equation 
(\ref{GPe}) in two different regimes: a crystal ($q=4$) case, and an 
irrational case ($T= \pi/ (\surd 5 +1)$). The evolution of the 
the kicked oscillator is conveniently studied by using a discretized 
propagator\cite{BR}:
\begin{equation}
\psi(x',t')\,=\, \int \, dx\, {\cal G}(x',x;t) \psi(x,0)
\end{equation}
where 
\begin{equation}
{\cal G}(x',x;t)\,=\, C \exp \left\{ \frac{i m_0 \omega_0}{2 \hbar \sin 
(\omega_0 t)} \left( (x^2+ x'^2)\cos (
\omega_0 t) -2 x x' \right) \right\}
\end{equation}
and once the discretized positions $x_i=(i-N/2)\Delta_x$ are introduced 
we have that the propagator ${\cal G}$ is unitary if we put
\begin{equation}
\Delta_x=\left( \frac{2 \pi \hbar \sin(\omega_0 t )}{m_0 \omega_0 N }
\right)^{1/2}
\end{equation}
In the coordinate representation the action of kicks is multiplicative 
on the wave function.
\begin{figure}[h]
\label{fig-2}
\centerline{\epsfxsize=5.5cm \epsfbox{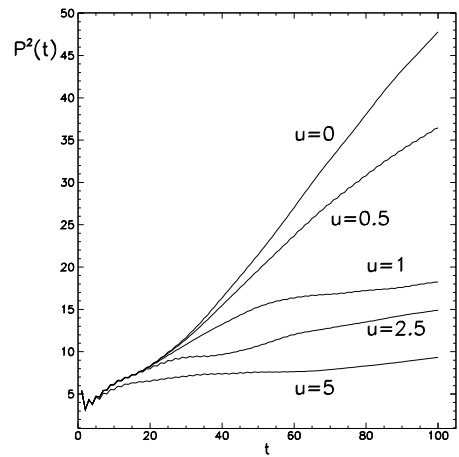}}
\caption{Quantum transport ($q=4$ one parameter symmetry group). See the text for details.}
\end{figure}
Simulation of quantum evolution is considerably more complicated once we introduce a 
nonzero nonlinearity: to propagate the wave function between kicks we 
separate the time independent part of the hamiltonian into the 
oscillator and the nonlinear part, and use the lowest order split 
method\cite{split} (this typically requires using ten time steps between 
consecutive kicks in order to get stable results).  As initial state we 
consider the ground state of the GP equation (without kicks) shifted in 
in the chaotic region nearest to the origin. The ground states for 
different values of the nonlinearity parameter are 
obtained by evolving an eigenstate of the quantum harmonic oscillator 
under the imaginary time Gross-Pitaevskii equation\cite{Baer}.

The first case we take into account is a {\em crystal} $q=4$ example.
In fig. ($1$) we 
show the classical phase picture, exhibiting unbounded transport along 
the stochastic web. 
We put $T=\pi/2$ (and $\omega_0$ fixed in such a way that generalized 
translations form a one parameter group of symmetries), $\epsilon=0.7$ (all others linear quantum 
parameters are fixed by taking $\hbar=1$ and $\xi=\surd 2$: we will 
adopt this choice for other examples too). 

This case is characterized by a one-parameter group of simmetries 
(generalized phase space translations) and thus the quantum case is 
expected\cite{BR} to 
exhibit a diffusive momentum spreading (the evolution corresponds of the upper 
line in fig. ($2$)): the effect of nonlinearity is considering 
by taking $u=0.5,\, 1,\,2.5$ and $5$: the corresponding curves are 
shown in fig. ($2$). Such simulations are 
performed by using a $N=2^{14}$ discretization of the position 
variable (which has been checked to provide a reliable choice up to the 
considered evolution time, by comparing the results with a simulation 
with twice the number of points). To smooth out oscillations in the 
evolving patterns we plot the integrated second moment
\begin{equation}
P^2(t)\,=\, {1 \over t}\sum_{k=0}^{t-1}\,\langle (p_k-p_0)^2 \rangle
\label{P-int}
\end{equation} 
\begin{figure}[h]
\label{fig-4d}
\centerline{\epsfxsize=5.5cm \epsfbox{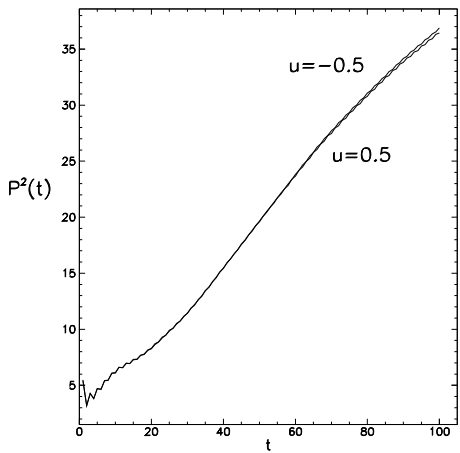}}
\caption{$P^2(t)$, for $u=+0.5$ and $u=-0.5$, and same
conditions as fig. ($2$).}
\end{figure}
The qualitative features confirm the 
observation in \cite{GJDCZ}, namely that the most striking effect of 
the nonlinearity is to oppose quantum delocalization. This has been 
claimed to be due to symmetry breaking effects of the nonlinearity, 
inhibiting transport along delocalized Floquet states. We remark that 
also at a classical level related features have been observed: if noise 
is added to the kicked Harper map, transport along the stochastic web 
is slowed down\cite{ClH}.
\begin{figure}[h]
\label{fig-3}
\centerline{\epsfxsize=5.5cm \epsfbox{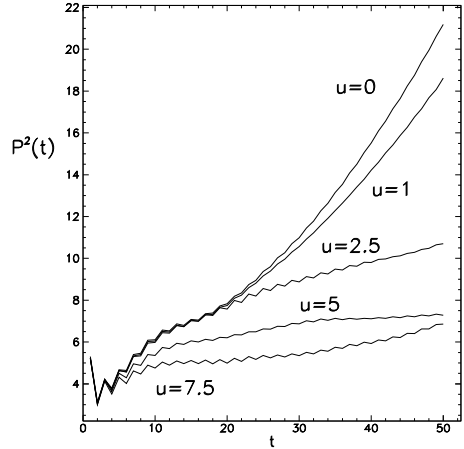}}
\caption{Quantum transport ($q=4$ two-parameter symmetry group) group, $\epsilon=0.7$. The configuration 
space has been discretized with $2^{12}$ points. 
}
\end{figure}

In principle a positive nonlinearity acts like a repulsive potential, 
but here the symmetry breaking effect is not related to the sign of the 
effective potential, as we see from fig.($3$), where it is shown how 
the sign of the nonlinearity has a tiny effect on momentum spreading.
\begin{figure}[h]
\label{fig-4}
\centerline{\epsfxsize=5.5cm \epsfbox{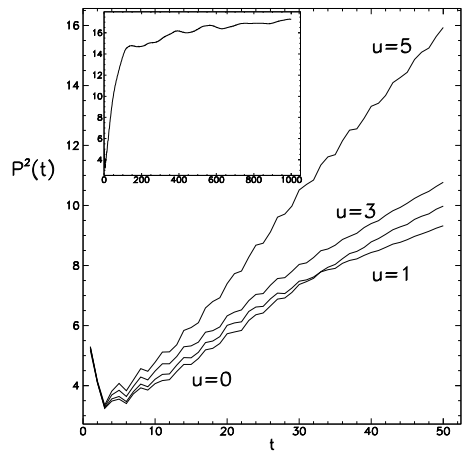}}
\caption{Quantum transport irrational case. Discretization is over 
$2^{14}$ points. The inset shows the kicked oscillator case on a longer 
time scale.}
\end{figure}
If such a view is correct, the effect must equally appear in the 
resonant crystal case, where a group of two-parameter symmetries leads 
to ballistic transport\cite{BR}: such an effect is indeed evident even 
for short times, see fig. ($4$): after a characteristic time 
scale, which shrinks as nonlinearity increases, the deviation from the 
kicked oscillator case is more and more marked as $u$ increases.

A priori the situation is different when we consider the oscillator 
outside the {\em crystal} regime: to this end we analyzed the case in 
which $\epsilon=1.4$ and $T= \pi/(\surd 5 +1)$. In this case the kicked 
oscillator displays dynamical localization: the striking observation is that here the 
nonlinearity acts in an opposite way, enhancing the quantum 
delocalization, see fig. ($5$). So, when symmetries are not 
present in the quantum case, nonlinearity seems to play a 
completely different role. This is at least in qualitative agreement 
to what happens to the kicked rotator evolving under a nonlinear 
Schroedinger equation, or even when noise is superimposed to the quantum
evolution\cite{OAH}. We 
have checked that the same happens even for higher values of 
$\epsilon$, when the oscillator undergoes a delocalization transition.

In conclusion, we have analyzed how nonlinearity influences a complex, and physically
relevant
quantum system, the kicked harmonic oscillator. We provide evidences 
that, at least at moderate times, it opposes quantum diffusion when 
transport is linked to symmetry properties of the linear hamiltonian, 
while it may lead to diffusion enhancement when no symmetry breaking 
occurs.

This work was partially supported by the PRIN-2000 project {\em Chaos and 
Localization in classical and quantum systems}, and the EU contract QTRANS
network (Quantum Transport on an Atomic Scale). We thank E. Arimondo and
J.H. M\"{u}ller for pointing ref. \cite{GJDCZ} to our attention.



\begin{thebibliography}{99}
\bibitem{note1} An asymptotic localized regime is also found for 
different classical control parameter values, where transport is of 
anomalous nature\cite{ArRu}.
\bibitem{KAMnote} This is possible as the model does not satisfy the 
requirements of Kolmogorov-Arnold-Moser theorem.
\bibitem{obsGP} We are considering wave functions normalized to unity, 
so that $u$ may be espressed as $4\pi N \hbar^2 a_s/m_0$, where $N$ is 
the number of particles in the condensate, and $a_s$ is the s-wave 
scattering length.
\end{thebibliography}
\end{document}